\documentclass[twocolumn,showpacs,preprintnumbers,amsmath,amssymb]{revtex4}

\usepackage{graphicx}
\usepackage{dcolumn}
\usepackage{bm}

\begin{document}

\title{Repulsive Casimir force between silicon dioxide and superconductor}

\author{Anh D. Phan}
\affiliation{Department of Physics, University of South Florida, Tampa, Florida 33620, USA}
\email{anhphan@mail.usf.edu}

\author{N. A. Viet}
\affiliation{Institute of Physics, 10 Daotan, Badinh, Hanoi, Vietnam}%

\date{\today} 

\begin{abstract}
We have presented a detailed investigation of the Casimir interaction between the superconductor $Bi_{2}Sr_{2}CaCu_{2}O_{8+\delta}$ (BSCCO) and silicon dioxide with the presence of bromobenzene in between. We found the dispersion force is repulsive and the magnitude of the force can be changed by varying the thickness of object and temperature. The repulsive force would provide a method to deal with the stiction problems and bring much meaningful from practical views.
\end{abstract}

\pacs{Valid PACS appear here}
\maketitle

The Casimir force, one of the most important causes of stiction problems, gives rise to critical impediments in fabrication and operation of nano/micro-electromechanical (NEMS/MEMS) systems. In almost all cases in which the scale is of hundreds of nanometers, the Casimir interactions produce such a significant amount of friction as to draw the much attention of scientists  from a wide variety of research fields \cite{bib1,bib2,bib3}. In addition, the theoretical understanding and measurements of the Casimir interactions have grown substantially in the last ten years, allowing physicists to have a more detailed understanding of fundamental physics, not only in nanophysics, but particle physics and cosmology as well. 

The simplest Casimir system was produced theoretically by Casimir in 1948 when he developed a model describing the interaction between two parallel conducting plates \cite{bib4}. Since then, the Casimir forces between real materials such as metals \cite{bib5}, semiconductors \cite{bib6}, semimetals \cite{bib7} and high-Tc superconductors \cite{bib8} have been extensively studied theoretically and experimentally. It has been found that the presence of liquids between two objects allow the sign of the Casimir force to switch \cite{bib1,bib2,bib3}. These repulsive Casimir forces appear when the dielectric functions of object 1 and object 2 immersed in a medium 3 satisfy the relation $\varepsilon_{1}(i\xi) < \varepsilon_{3}(i\xi) < \varepsilon_{2}(i\xi)$ over a wide imaginary frequency range $\xi$. It is also possible to make the repulsion using arrays of gold nanopillars on two plates \cite{bib9}. In addition, metamaterials are promising candidates for creating these repulsive interactions \cite{bib10}. The combination between experimental measurements and theoretical calculations have provided essential information to help researchers design nanoscale devices. 

The well-known Lifshitz theory developed the generalization of the Casimir force \cite{bib3,bib11,bib12}. In the theory, the force between uncharged objects made of real materials was given by an analytical formula with the frequency-dependent dielectric permittivity $\varepsilon(i\xi)$ . The variation of the dielectric functions causes the change of the Casimir interactions. The Casimir-Lifshitz force has been studied at for systems at the thermal equilibrium between the atom-atom, plane-plane and atom-plane configurations \cite{bib11,bib20}. 

The cuprate superconductor, the high-Tc superconductor and the anisotropic materials are widely used in various devices. It has been shown theoretically that the Casimir force in the BSCCO-air-gold system is significantly affected by the anisotropy in the dielectric functions. In the present paper, the Casimir-Lifshitz force is calculated in the case of the perpendicular cleave between the cuprate superconductor and silica with bromobenzene in between. The force calculation take into account the thermal effect and the influence of its thickness on the dispersion force. The force is repulsive and deals with the sticking process in nano devices.

The general expression describing the Casimir interaction between two infinite parallel plates is the Lifshitz formula. At a given separation $a$ and given temperature $T$, the Casimir pressure between two plates is given \cite{bib11,bib12}
\begin{eqnarray}
P(a)=-\dfrac{k_{B}T}{\pi}\sum_{n=0}^{\infty} \int_{0}^{\infty}qk_{\perp}dk_{\perp} \sum_{\alpha}\dfrac{r_{\alpha}^{(1)}r_{\alpha}^{(2)}}{e^{2qa}-r_{\alpha}^{(1)}r_{\alpha}^{(2)}},
\label{eq:1}
\end{eqnarray}
here $k_{B}$ is the Boltzmann constant, $k_{\perp}$ is the wave vector component perpendicular to the plate, $\alpha = TM,TE$, $r_{TM}^{(1,2)}$ and $r_{TE}^{(1,2)}$ denote the reflection coefficients of the transverse magnetic (TM) and transverse electric (TE) field, respectively. The superscript (1) and (2) correspond to the first body (silica) and the second body (BSCCO). In addition, $q = \sqrt{k_{\perp}^{2} + \varepsilon_{3}\xi_{n}^{2}/c^{2}}$, $\xi_{n}=2\pi nk_{B}T/\hbar$ are the Matsubara frequencies,$n$ is an integer, and $\varepsilon_{3}\equiv\varepsilon_{3}(i\xi_{n})$ is the dielectric function of medium in between two objects. In the calculation, bromobenzene is medium. The dielectric function of the liquid can be described using the oscillator model \cite{bib2}
\begin{eqnarray}
\varepsilon_{3}(i\xi) = 1 + \sum_{i}\frac{C_{i}}{1+\xi^2 /\omega_{i}^2},
\label{eq:2}
\end{eqnarray}
where parameters $C_{i}$ and $\omega_{i}$ were obtained by fitting with experimental data in the large range of frequency \cite{bib2}. 

It is important to note that for $n=0$, the prefactor of the integration is $k_{B}T/(2\pi)$ instead of $k_{B}T/\pi$ for other values of $n$. In the case of silicon dioxide, the reflection coefficients are presented \cite{bib2,bib3}
\begin{eqnarray}
r_{TM}^{(1)}=\frac{\varepsilon_{1}q-\varepsilon_{3} \sqrt{k_{\perp}^{2} + \varepsilon_{1}\xi_{n}^{2}/c^{2}}}{\varepsilon_{1}q+\varepsilon_{3} \sqrt{k_{\perp}^{2} + \varepsilon_{1}\xi_{n}^{2}/c^{2}}},
\label{eq:3}
\\
r_{TE}^{(1)}=\frac{q-\sqrt{k_{\perp}^{2} + \varepsilon_{1}\xi_{n}^{2}/c^{2}}}{q+\sqrt{k_{\perp}^{2} + \varepsilon_{1}\xi_{n}^{2}/c^{2}}},
\label{eq:4}
\end{eqnarray}
in which $\varepsilon_{1}\equiv\varepsilon_{1}(i\xi_{n})$ is the dielectric function of silica. The dielectric fuction still has the form of an oscillator model as in Eq.(\ref{eq:2}) and the parameters were generated in Ref.\cite{bib2}. Considering the role of the thickness $D$ of the silica slab on the Casimir interaction, the reflection coefficients $TM$ and $TE$ in Eq.(\ref{eq:3}) and Eq.(\ref{eq:4}) become \cite{bib13,bib14}
\begin{eqnarray}
r_{TM}^{(1)}\rightarrow r_{TM}^{(1)}\frac{1-e^{-2\sqrt{k_{\perp}^{2} + \varepsilon_{1}\xi_{n}^{2}/c^{2}}D}}{1-r_{TM}^{(1)2}e^{-2\sqrt{k_{\perp}^{2} + \varepsilon_{1}\xi_{n}^{2}/c^{2}}D}},
\label{eq:5}
\\
r_{TE}^{(1)}\rightarrow r_{TE}^{(1)}\frac{1-e^{-2\sqrt{k_{\perp}^{2} + \varepsilon_{1}\xi_{n}^{2}/c^{2}}D}}{1-r_{TE}^{(1)2}e^{-2\sqrt{k_{\perp}^{2} + \varepsilon_{1}\xi_{n}^{2}/c^{2}}D}}.
\label{eq:6}
\end{eqnarray}

Because of the uniaxial property of BSCCO, its permittivity in the perpendicular cleave is represented in the form of a tensor \cite{bib8}
\begin{eqnarray}
\Bar{\varepsilon_{2}} = \left[ \begin{array}{ccc} \varepsilon_{2\perp} & 0 & 0 \\
 0 & \varepsilon_{2\perp} & 0 \\
 0 & 0 & \varepsilon_{2||}\end{array} \right]
\label{eq:7}
\end{eqnarray}
where $\varepsilon_{2||}$ and $\varepsilon_{2\perp}$ are the dielectric components along the optical axis and perpendicular to the optical axis. Therefore, the expression of the reflection coefficients for BSCCO must be modified. The TM and TE coefficients on the liquid-BSCCO interface are performed \cite{bib8,bib15}
\begin{eqnarray}
r_{TM}^{(2)}=\frac{\varepsilon_{2\perp}q-\varepsilon_{3} \sqrt{\frac{\varepsilon_{2\perp}}{\varepsilon_{2||}} k_{\perp}^{2} + \varepsilon_{2\perp}\xi_{n}^{2}/c^{2}}}{\varepsilon_{2\perp}q+\varepsilon_{3} \sqrt{\frac{\varepsilon_{2\perp}}{\varepsilon_{2||}} k_{\perp}^{2} + \varepsilon_{2\perp}\xi_{n}^{2}/c^{2}}},,
\label{eq:8}
\\
r_{TE}^{(2)}=\frac{q-\sqrt{k_{\perp}^{2} + \varepsilon_{2\perp}\xi_{n}^{2}/c^{2}}}{q+\sqrt{k_{\perp}^{2} + \varepsilon_{2\perp}\xi_{n}^{2}/c^{2}}}.
\label{eq:9}
\end{eqnarray}

For BSCCO, the dielectric response $\varepsilon_{2||}$ and $\varepsilon_{2\perp}$ are modeled based on the damped-multioscillator model. Parameters corresponding to the resonance frequency, damping and oscillator strength were given in Ref.\cite{bib8}.

\begin{figure}[htp]
\includegraphics[width=9.5cm]{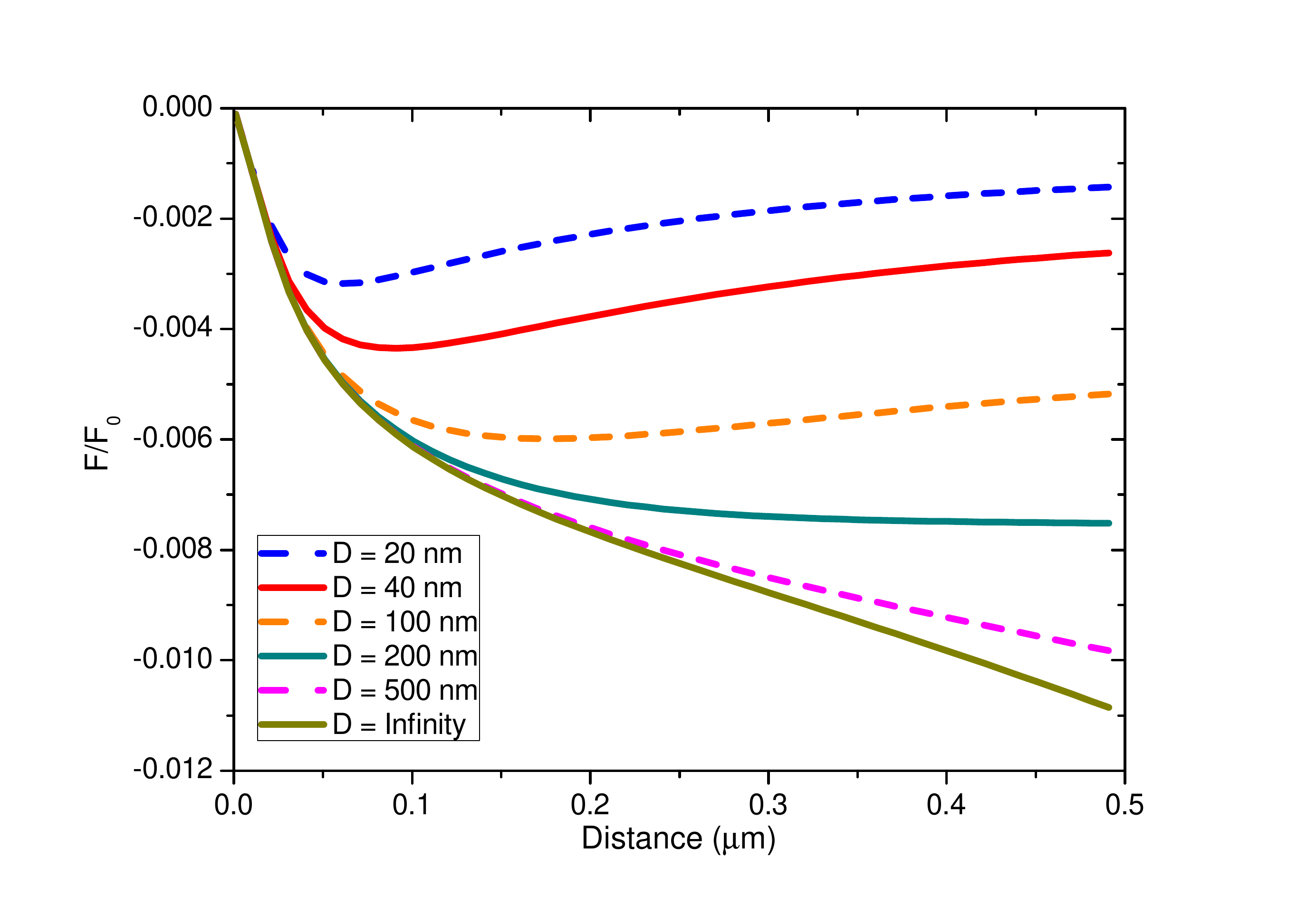}
\caption{\label{fig:1}(Color online) The relative Casimir pressures between a BSCCO plate and a silica slab 
in the presence of bromobenzene, here $F_{0}(a)=\pi^2\hbar c/(240a^4)$ is the Casimir force between two ideal metal plates.}
\end{figure}

As shown in Fig.~\ref{fig:1}, the Casimir forces are significantly influenced by the thickness of the slab. The Casimir interactions in the real system are much smaller than the force in the ideal case. It is clear that the presence of bromodenzene makes the Casimir force in this case is repulsive. For $D>500$ $nm$, the effect of thickness nearly vanishes and the Casimir force in the system can be modeled as an interactions between two plates. For metal such as gold, when the thickness $D > 30$ $nm$, a gold thin film can be treated as a gold plate \cite{bib16,bib17}. The reason for the discrepancy between the two cases is that the conductivity of metal is much larger than that of silicon dioxide. When the thickness of a metal slab is reduced, it appears as if the skin-depth effect occurs. This effect, however, is not present in the case of silica. 

At small distances, there is not much change in the force with different thicknesses. The reason is that at this range $D/a >> 1$, so $e^{-2\sqrt{k_{\perp}^{2} + \varepsilon_{1}\xi_{n}^{2}/c^{2}}D} << 1$. The influence of thickness on the interaction disappears. 

\begin{figure}[htp]
\includegraphics[width=9.5cm]{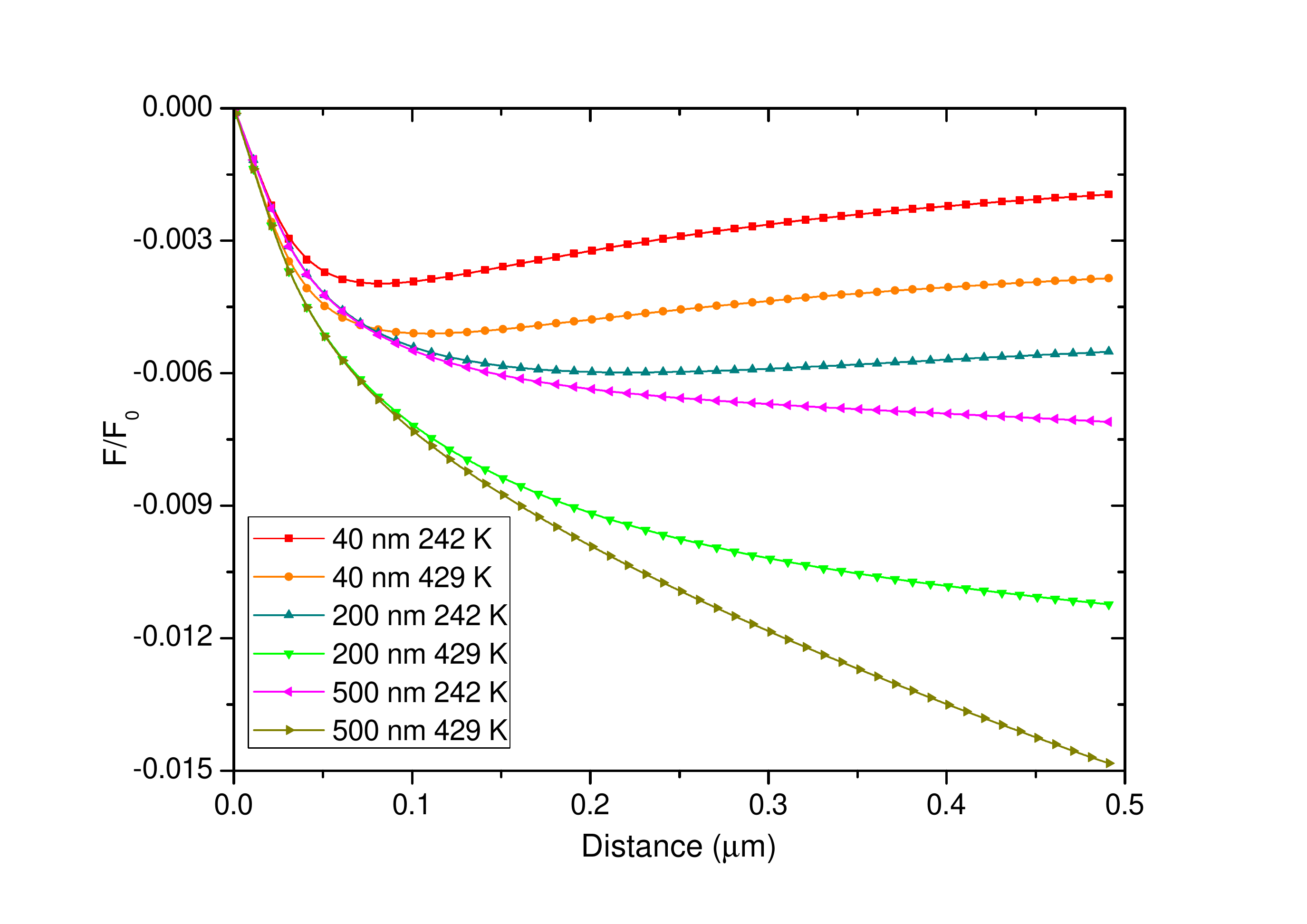}
\caption{\label{fig:2}(Color online) The relative Casimir pressures are taken into account the thermal effect with different values of thickness.}
\end{figure}

Bromobenzene molecules exist in liquid form over an important range of temperature from $242$ $K$ to $429$ $K$. Fig.~\ref{fig:2} shows the Casimir force in this temperature range. It is evident that the interaction depends notably on temperature. There are variations in the Casimir force at different thicknesses. It is now possible to design a non-touching system because the Casimir force is repulsive for entire range of distance. The Casimir force, and the gravitational forces between the two bodies, with the earth lead to the repulsive-attractive transition in our system and results in our system reaching an equilibrium distance \cite{bib18}. Obviously, the stable position can be varied by changing the sizes of bodies, the thickness and temperature. In Ref.\cite{bib19}, authors presented the proposal for measuring the Casimir force with the presence of a tiny spring in order to get a balanced position and the oscillation frequency. However, in this case, the equilibrium positions exist naturely. It is not necessary to attach a spring to the system to measure the Casimir force. The force can be found via observation of the oscillation frequencies. Because of the anisotropic property, the expressions of the reflection coefficients TE and TM in the case of the parallel cleave orientation are different from Eq.(\ref{eq:8}) and Eq.(\ref{eq:9}). This discrepancy is proof that there is a difference between the Casimir forces in two orientations.

The thermal effect in the Casimir interaction plays an important role at long distances \cite{bib12}. For short distances, this effect can be ignored. One can used the double integration instead of summation and single integration as Eq.(\ref{eq:1}) in calculations at short distances. The expression of the double integration provides a good agreement with experiment. 

This work presents a reliable anti-stiction method of NEMS/MEMS structures. The presence of liquid can address the stiction issue causing catastrophic failure in nanoscale devices. The temperature and thickness depedence of the Casimir force allows control of the adhesion force between two surfaces. It is currently difficult to measure properties in fluidic environments. However, using liquid films in NEMS/MEMS devices with the range of the thickness of liquid layer from 2 $nm$ to 70 $nm$ has been intensively investigated \cite{bib21,bib22}. Moreover, authors in \cite{bib23} described the behavior of the tiny devices in liquids. These research makes it possible to design nanostructures in the microfluidic environment and our studies give an interesting view of what happens physically in systems submerged in liquids.

\begin{acknowledgments}
The work was partly funded by the Nafosted Grant No. 103.06-2011.51.
\end{acknowledgments}

\newpage 

\end{document}